# ESTIMATION OF THE LOWEST LIMIT OF 1/f NOISE IN SEMICONDUCTOR MATERIALS

FERDINAND GRÜNEIS

*Institute for Applied Stochastic*
*Rudolf von Scholtz Str. 4, 94036 Passau, Germany*
*Email: Ferdinand.Grueneis@t-online.de*

A lowest limit of 1/f noise in semiconductor materials has not yet been reported; we do not even know if such a lowest limit exists. 1/f noise in semiconductors has recently been brought into relation with 1/f noise in quantum dots and other materials. These materials exhibit on-off states which are power-law distributed over a wide range of timescales. We transfer such findings to semiconductors, assuming that the g-r process is also controlled by such on-off states. As a result, we obtain 1/f noise which can be expressed as Hooge's relation. Based on the intermittent g-r process, we estimate the lowest limit of 1/f noise in semiconductor materials. We show that this limit is inversely proportional to the dopant concentration; to detect the lowest limit of 1/f noise, the number of centers should be as small as possible. We also find a smooth dependence of 1/f and g-r noise on time.

## 1. Introduction

For a homogenous semiconductor and resistor, Hooge [1] empirically found

$$\frac{S_{1/f}(f)}{I_0^2} = \frac{\alpha}{N_0}\frac{1}{f}. \tag{1}$$

$S_{1/f}(f)$ is the power spectral density of a fluctuating current $I(t)$ with a mean value of $I_0$ and $N_0$ is the number of charge carriers in the probe volume. $\alpha$ is the so-called Hooge coefficient; empirically $10^{-6} < \alpha < 10^{-3}$ has been found [2-4]. Most researchers agree that 1/f noise in semiconductors and metallic resistors is well described by Eq. (1). However, several problems arise with Eq. (1):

- the total power of 1/f noise in Eq. (1) is proportional to the integral

$$\int_{f_l}^{f_u}\frac{df}{f} = ln\left(\frac{f_u}{f_l}\right) \tag{2}$$

  where $f_l$ is a lower and $f_u$ an upper frequency limit of 1/f noise. If $\alpha$ is a constant independent of time, the integral diverges if $f_l$ tends towards zero. Caloyannides observed a 1/f shape down to a frequency of $10^{-6.3}$ Hz without finding a lowest limit [5]. The 1/f shape is assumed to continue to even lower frequencies; it is only the observation time and the stability of recording instruments which sets a limit to $f_l$. It is an open question whether such a lowest limit exits at all.
- Hooge's relation applies to a pure 1/f shape; however, in most cases a $1/f^b$ shape with a slope $0.8 < b < 1.2$ is observed. This makes it difficult to compare the Hooge coefficient $\alpha$ of different materials exhibiting a slope $b \neq 1$.
- A generally accepted physical interpretation of the Hooge coefficient $\alpha$ does not yet exist.

1/f noise in semiconductors has recently been brought into relation with 1/f noise observed in quantum dots, nanowires and some organic molecules [6-8]. These materials exhibit periods of bright ("on") and dark ("off") states; this phenomenon is characterized by fluorescence intermittency. These on-off states follow power-law statistics over a wide range of timescales from micro-seconds to minutes. These findings have been transferred to semiconductor materials when applying intermittency to the non-radiative g-r process [9]. As a result, a generalized form of Hooge's relation has been derived which also applies also for a slope $b \neq 1$. The predicted Hooge coefficient $\alpha_H$ was shown to depend on the parameters of g-r noise and on time.

In this paper, we address the problem of the lowest limit of 1/f noise. The intermittent g-r process exhibiting power-law distributed on-times gives rise to 1/f noise. We estimate the probability for the longest on-time and the corresponding lowest frequency limit. The paper is organized as follows: Section 2 presents the essential features of the intermittent g-r process. The results of this section have already been published in [9] and are summarized as far as they are of relevance for the following. In Section 3, we estimate the lowest frequency limit to 1/f noise and investigate the dependence of 1/f noise and g-r noise on time.

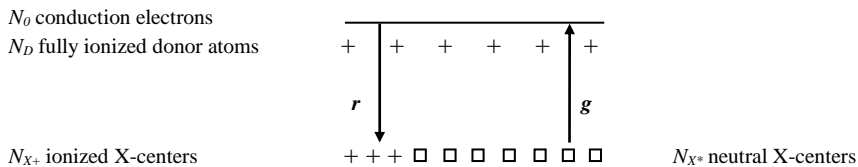

Fig. 1. The energy band system and electron transitions involved in the fluctuations of conduction electrons in a two-level system. $N_X$ is the number of X-centers, $N_D$ is the number of fully ionized shallow donor atoms. Under steady state conditions, $N_{X^+}$ is the number of ionized X-centers, and $N_{X^*}$ is the number of neutral X-centers. $N_0$ is the number of conduction electrons.

## 2. Fluctuations in a Semiconductor

This paper's considerations are confined to g-r noise in a two-level system of a doped n-type semiconductor. Thermal noise that is always present is not discussed. We consider only transitions between the level of X-centers and the conduction bands as producing fluctuations (see Fig. 1). For example, X-centers may be donor or trap atoms. The number of X-centers is denoted by $N_X$ and the number of fully ionized shallow donor atoms by $N_D$. The mean number of ionized X-centers is $N_{X+}$, of neutral X-centers is $N_{X*} = N_X - N_{X+}$ and that of conduction electrons is $N_0 = N_{X+} + N_D$.

### 2.1. *The master equation approach to g-r noise*

Under steady state conditions, the master equation approach leads to the following expressions [10-13]

$$g_0 = \overline{\Delta N^2} / \tau_{gr} \tag{3}$$

$$S_{gr}(f) = \left(\frac{I_0}{N_0}\right)^2 \frac{4 g_0 \tau_{gr}^2}{1+(2\pi f \tau_{gr})^2}. \tag{4}$$

$g_0$ is the generation rate, $\overline{\Delta N^2}$ is the mean square fluctuations of conduction electrons about the mean value of conduction electrons $N_0$ and $\tau_{gr}$ is the relaxation time of the g-r process. $S_{gr}(f)$ is the power spectral density of the current fluctuations resulting from an applied current $I_0$.

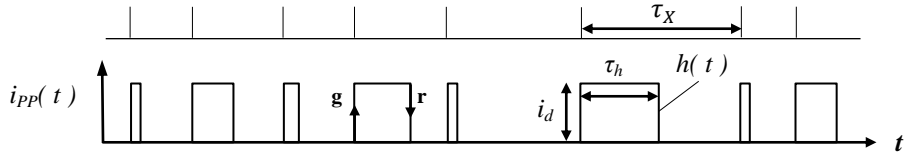

Fig. 2. Top: The spike train indicates the time points for the generation of an electron at a single X-center; the time between the successive generations is $\tau_X$. Bottom: Current fluctuations due to a single X-center. The time series $i_{PP}(t)$ consists of successive rectangular current pulses $h(t)$ with lifetime $\tau_h$ and amplitude $i_d$. The pulse $h(t)$ represents an elementary g-r process. After generation at a single X-center, an electron remains for the lifetime $\tau_h$ in the conduction band before it recombines to an arbitrary ionized X-center. A mean drift current is $i_d = e\mu E_0/L$, where $e$ is the elementary charge, $\mu$ is the mobility, $E_0$ is an applied electric field, and $L$ is the length of the sample.

### 2.2. *The shot noise interpretation of g-r noise*

To compare with <u>single</u> quantum dots and other materials, the noise contribution due to a <u>single</u> X-center is considered; this is defined by

$$S_X(f) = \frac{S_{gr}(f)}{N_X}. \tag{5}$$

Correspondingly, the generation rate due to a single X-center is

$$\frac{1}{\tau_X} = \frac{g_0}{N_X}, \tag{6}$$

where $\tau_X$ is the time between successive generations (see the top of Figs. 2 and 3). As shown at the bottom of Fig. 2, each spike triggers an elementary g-r pulse $h(t)$ leading to a time series $i_{PP}(t)$. This random succession of g-r pulses can be interpreted as shot noise with power spectral density [14]

$$S_X(f) = \frac{2}{\tau_X} \overline{|H(f)|^2}. \tag{7}$$

$H(f)$ is the Fourier transform of $h(t)$. As is well-known [11-13], the lifetime $\tau_h$ of conduction electrons is exponentially distributed with mean value $\tau_{gr}$ yielding

$$\overline{|H(f)|^2} = \frac{2 i_d^2 \tau_{gr}^2}{1+(2\pi f \tau_{gr})^2}. \tag{8}$$

The noise contribution due to the $N_X$ centers is $N_X S_X(f)$. Taking into account Eq. (6) and considering that a current $I_0 = N_0 i_d$ we end up with Eq. (4). Therefore, the shot noise interpretation of g-r noise is consistent with the master equation approach to the g-r noise in Eq. (4).

### 2.3. *An intermittent g-r process as a possible origin of 1/f noise*

As mentioned in the Introduction, we transfer the phenomenon of intermittency to semiconductor materials in assuming that the g-r process is also controlled by "on-off" states. By analogy to a blinking <u>single</u> quantum dot, we regard a <u>single</u> X-center. The g-r process of such a single X-center is assumed to be intermitted by a gating function with $\tau_{off}$ being the off-time; the on-time is denoted by $\tau_{on}$ (see Fig. 3.b). In this way the Intermittent Poisson Process (= IPP) is obtained as is shown in Fig. 3.c. Each spike triggers a g-r pulse $h(t)$ as is seen in Fig. 2 giving rise to an intermittent g-r process.

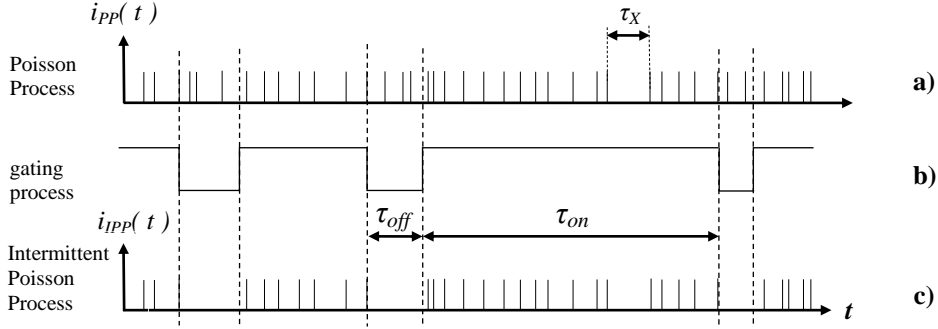

Fig. 3.a. Schematic plot of a Poisson Process (= PP) with time series $i_{PP}(t)$ representing the occurrence of successive g-r pulses due to a single X-center; $\tau_X$ is the time between successive g-r pulses. Fig. 3.b. The Poisson Process is gated by a two-state process with states $\tau_{off}$ (= intermission) and $\tau_{on}$ (= duration of a cluster). Fig. 3.c. The Intermittent Poisson Process (= IPP) is characterized by intermissions followed by fluctuating clusters. Each spike triggers a g-r pulse $h(t)$ as is seen in Fig. 2.

As in quantum dots and other materials, the on-times are assumed to be power-law distributed such as $1/t^{\mu_{on}}$. This leads to a finite and random number of spikes to so-called clusters (see Fig. 3.c). These number fluctuations can be described by a cluster size distribution $q_n$ that follows the power-law distribution of on-times $n = 1, 2, \ldots M_t$:

$$q_n \propto 1/n^{\mu_{on}}. \tag{9}$$

$q_n$ is a truncated Zeta or Zipf distribution [15]. $M_t$ is the maximum number of spikes in a cluster. Similar to blinking of single quantum dots and other materials the exponent $\mu_{on} \leq 2$ (see Section 2.6). The off-times are assumed to be exponentially distributed.

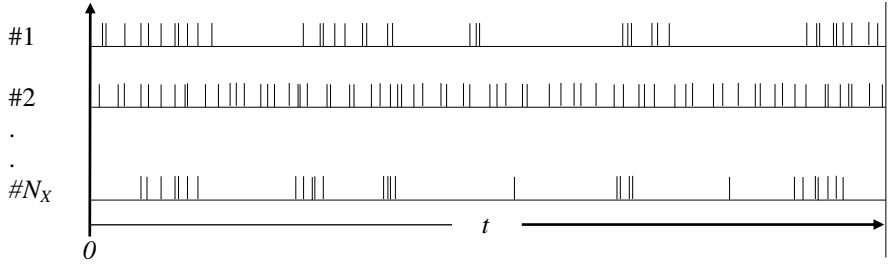

Fig. 4. The $N_X$ centers starting simultaneously at $t = 0$ with the generation of the IPP; each spike triggers an elementary g-r current pulse $h(t)$ as shown in Fig. 2. The time between spikes in a cluster is $\tau_X$. The X-centers are numbered #1, #2, ….#$N_X$. The X-center #2 shows a cluster without any intermissions during time $t$.

We define $t = 0$ as the time when the $N_X$ centers simultaneously start generating an Intermittent Poisson Process (= IPP); for an illustration see Fig. 4. Consequently, the time $t$ can be interpreted as the age of the semiconductor material after doping; at $t = 0$ the material is assumed to be in thermal equilibrium. Among the $N_X$ centers there will be a certain number of clusters without any intermissions during $t$ (see X-center #2 in Fig. 4). In such clusters, the mean number of events occurring within $t$ is $M_t$; this obeys Poisson statistics leading to[1]

$$M_t = \frac{t}{\tau_X}. \tag{10}$$

All of the quantities containing $M_t$ depend on time. Denoting the mean number of spikes in a cluster by $\overline{N_c} = \sum n\, q_n$ the mean on-time (= the mean duration of a cluster) is obtained by

$$\tau_{on} = \overline{N_c}\tau_X. \tag{11}$$

A mean cluster size $\overline{N_c}$ as a function of $\mu_{on}$ and $M_t$ is found in the Appendix.

## 2.2. 1/f noise and g-r noise in the probe volume

The Intermittent Poisson Processes $i_{IPP}(t)$ in Fig. 3.c represents the current fluctuations due to a single X-center. The total noise in the probe volume is obtained by [9]

$$S_{tot}(f) = \beta_{im} S_{gr}(f) + S_{1/f}(f). \tag{12}$$

The first term on the r.h.s. is reduced g-r noise containing a pre-factor

$$\beta_{im} = \frac{\tau_{on}}{\tau_{on}+\tau_{off}}. \tag{13}$$

The second term in Eq. (12) is 1/f noise which can be expressed by

$$S_{1/f}(f) = \frac{1}{2} \frac{\tau_X}{\tau_{on}+\tau_{off}} S_{gr}(f)\, \Phi_{1/f}(f). \qquad (14)$$

---

[1] In Eq. (9) $M_t$ is a step function. In Eq. (10) the step function can be interpolated yielding a function continuous in time $t$.

The spectral function $\Phi_{1/f}(f)$ can be expressed by

$$\Phi_{1/f}(f) \approx \frac{C(r)}{(f\tau_X)^b}, \qquad (15)$$

whereby

$$C(r) \approx 0.3 \left(\frac{r}{3+r}\right)^2 \qquad (16)$$

is a pre-factor with

$$r = \frac{\tau_{off}}{\tau_X} \qquad (17)$$

being the normalized off-time. For $r \ll 1$, the pre-factor $C(r) \approx 0.03\, r^2$, whereas for $r \gg 1$ it is $C(r) \approx 0.3$. A plausible explanation for this behavior is provided by the following consideration: for $r \gg 1$, the impact of intermission can be neglected, and 1/f noise is essentially determined by fluctuating on-times. For $r \ll 1$, however, the intermission is much smaller than the inter event time; eventually for $r \to 0$, the IPP degenerates to a Poisson process. Correspondingly, 1/f noise decreases rapidly if $r \ll 1$. The spectral function $\Phi_{1/f}(f)$ scales within the lower and upper cut-off frequency

$$f_l \approx 1/2 M_t \tau_X \qquad (18)$$

and

$$f_u \approx 1/2\pi\tau_X. \qquad (19)$$

Beyond $f_u$ the 1/f shape rapidly approaches zero; below $f_l$ the 1/f noise reaches a plateau. Hence, the scaling region of 1/f noise is

$$M_t \approx f_u/f_l \qquad (20)$$

An illustration of reduced g-r noise and 1/f noise and of the corresponding cut-off frequencies is shown in Fig. 5.

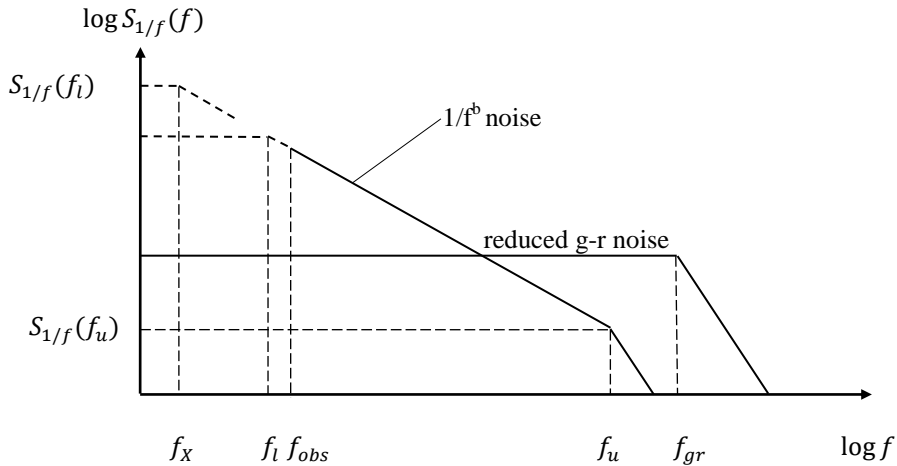

Fig. 5. Schematic plot of $1/f^b$ noise component $S_{1/f}(f)$. $f_l$ and $f_u$ are the lower and upper cut-off frequencies respectively. The upper frequency limit of reduced g-r noise $f_{gr} = 1/2\pi\tau_{gr}$ is well above $f_u$. The narrowest possible frequency resolution is $f_{obs}$; the lowest frequency limit to 1/f noise is $f_X$ (see Section 3).

### 2.4. *An alternative form of Hooge's relation*

Using Eqs. (4) and (14), the 1/f noise can be given a generalized form of Hooge's relation

$$\frac{S_{1/f}(f)}{I_0^2} = \frac{\alpha_H}{N_0} \frac{\tau_X}{(f\tau_X)^b}. \qquad (21)$$

For $b \to 1$, this reduces to Hooge's relation in Eq. (1). The predicted Hooge coefficient is

$$\alpha_H = \alpha_{im} \alpha_{gr}. \qquad (22)$$

Herein, the coefficient

$$\alpha_{im} = 2C(r) \frac{\tau_X}{\tau_{on}+\tau_{off}} \qquad (23)$$

depends on the parameters of the on-off intermittency and the coefficient

$$\alpha_{gr} = \frac{\overline{\Delta N^2}}{N_X} \frac{\overline{\Delta N^2}}{N_0} \qquad (24)$$

on the parameters of the g-r process. Dividing this by $N_0$ yields

$$\frac{\alpha_{gr}}{N_0} = \frac{1}{N_X}\left(\frac{\overline{\Delta N^2}}{N_0}\right)^2 \qquad (25)$$

transforming Eq. (21) to

$$\frac{S_{1/f}(f)}{I_0^2} = \frac{\gamma_H}{N_X}\frac{\tau_X}{(f\tau_X)^b}. \qquad (26)$$

Herein

$$\gamma_H = \alpha_{im}\left(\frac{\overline{\Delta N^2}}{N_0}\right)^2 \qquad (27)$$

is the alternative Hooge coefficient relating 1/f noise to $N_X$ (= the number of X-centers) rather than to $N_0$ (= the number of charge carriers) as is defined by Hooge in Eq. (1).

### 2.6. *Extreme property of the slope b*

The slope $b$ in Eq. (26) is determined by (see Fig. 5)

$$b = -\frac{\log S_{1/f}(f_l) - \log S_{1/f}(f_u)}{\log f_l - \log f_u}. \qquad (28)$$

Taking Eq. (14) into account and considering also that $f_u < f_{gr}$ this is equivalent to

$$b = -\frac{\log \Phi_{1/f}(f_l) - \log \Phi_{1/f}(f_u)}{\log f_l - \log f_u}. \qquad (29)$$

At $f_l$ the spectral function $\Phi_X$ is calculated by [16]

$$\Phi_X(f_l) \approx \Phi_X(0) \approx \overline{N_c^2}\left(\frac{r}{r+\overline{N_c}}\right)^2. \qquad (30)$$

At $f_u$ the spectral function $\Phi_X$ is for

$r \leq 1$:
$$\Phi_X(f_u) \approx 2r^2(q_1 - q_2 - q_1^2) \qquad (31)$$

and

$r > 1$:
$$\Phi_X(f_u) \approx 2(1 - q_1). \qquad (32)$$

Substituting this into Eq. (29), we distinguish two cases applying for $M_t \gg 1$ and for

$r \leq 1$:
$$b \approx \log\left\{\frac{\overline{N_c^2}}{\overline{N_c^2}}\right\}/\log M_t \qquad (33)$$

and

$r > 1$:
$$b \approx \log\left\{\overline{N_c^2}\left(\frac{r}{r+\overline{N_c}}\right)^2\right\}/\log M_t. \qquad (34)$$

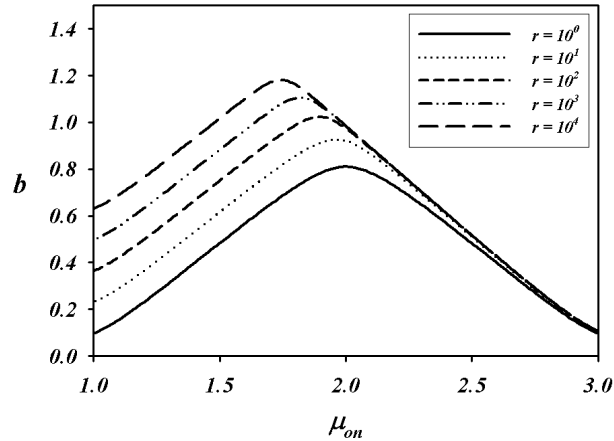

Fig. 6. Slope $b$ according to Eq. (34) as a function of exponent $\mu_{on}$ for several values of $r$ and for $M_t = 10^{15}$. The curves for $r \leq 1$ coincide with the curve for $r = 1$.

Based on Eqs. (33) and (34), Fig. 6 shows slope $b$ as a function of exponent $\mu_{on}$ for several values of $r$ and for a scaling region $M_t = 10^{15}$. Depending on the normalized off-time $r$, the slope exhibits a distinct maximum between $0.8 < b < 1.2$. This maximum is found for $1.75 < \mu_{on} \leq 2$.

We adapt the parameter $\mu_{on}$ to this extreme property[2]; under this condition, the choice of the exponent $\mu_{on}$ is determined by the maximum value of the slope $b$. For the following, we denote the maximum value of the slope by $\hat{b}$ and the corresponding exponent by $\hat{\mu}_{on}$ (see Fig. 7). As is shown in [9], an rough estimate for $\hat{b}$ and for $\hat{\mu}_{on}$ is provided by the intersection of the two straight lines

$$b_1 = 3 - \mu_{on} \quad \text{and} \quad b_2 = \mu_{on} - 1 + 2\frac{\log r}{\log M_t}. \tag{35}$$

As an example, see Fig. 7, where these approximations are shown in comparison to slope $b$. Table 1 summarizes these approximations applying the better the larger $t$. Further variables deriving from $\hat{b}$ and $\hat{\mu}_{on}$ will be indexed by a head cap over the corresponding variable.

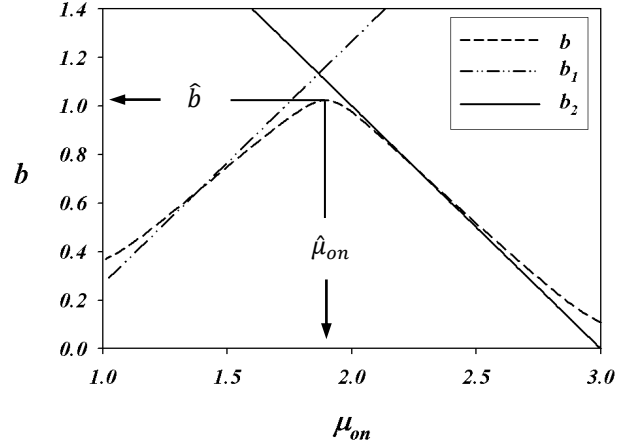

Fig. 7. Slope $b$ for $r = 10^2$ and $M_t = 10^{15}$ as a function of the exponent $\mu_{on}$ together with an approximate treatment by two straight lines $b_1$ and $b_2$. The maximum value of slope $b$ is denoted by $\hat{b}$ and the corresponding exponent by $\hat{\mu}_{on}$. The intersection of the two straight lines $b_1$ and $b_2$ provides a rough estimate for $\hat{b}$ and $\hat{\mu}_{on}$.

Table 1. Approximations for $\hat{b}(t)$ and $\hat{\mu}_{on}(t)$.

| | | | |
|---|---|---|---|
| $r \leq 1$ | $\tau_{off} \leq \tau_X$ | $\hat{b}(t) \approx 1 - \frac{2 \log (\ln M_t)}{\log M_t}$ | $\hat{\mu}_{on} = 2$ |
| $r > 1$ | $\tau_{off} > \tau_X$ | $\hat{b}(t) \approx 1 + \frac{\log r}{\log M_t}$ | $\hat{\mu}_{on}(t) \approx 2 - \frac{\log r}{\log M_t}$ |

## 3. The estimation of the lowest limit of 1/f noise in semiconductor materials

The Intermittent Poisson Process (= IPP) is the point process underlying the intermittent g-r process. Based on the IPP we provide a lowest limit to 1/f noise in semiconductor materials. Let's assume that an observation time starts at $t = 0$ (see Fig. 3). Then $T_{obs} = t$ and the narrowest possible frequency resolution is [20]

$$f_{obs} \approx 1/t. \tag{36}$$

We observe a 1/f shape which abruptly ends at frequency $f_{obs}$. According to Eq. (18), the lower limit to 1/f noise is

$$f_l \approx 1/2M_t\tau_X. \tag{37}$$

As long as $M_t$ increases with time, Eq. (10) applies leading to

$$f_l \approx 1/2t. \tag{38}$$

As time $t$ increases, both $f_l$ and $f_{obs}$ shift down in parallel to lower frequencies. However, $f_l$ being equal to ½ $f_{obs}$ remains unobservable (see also Fig. 5).

For convenience we denote a cluster containing $M_t$ events an "$M_t$-cluster" (as an example, see X-center #2 in Fig. 3). According to Eq. (A.5), the time-dependent probability for an $M_t$-cluster is

$$\hat{q}_{M_t} = \frac{\zeta(\hat{\mu}_{on})}{M_t^{\hat{\mu}_{on}(t)}}. \tag{39}$$

---

[2] Such an approach may be justified by the following considerations: the extreme value of slope $b$ is accompanied by an extreme variance of counts in time $M_t$ [16]. Shockley [17] showed that the g-r process is phonon-induced; this mechanism also applies to the intermittent g-r process. Hence, the intermittent g-r process as is seen in Fig. 3c, is eventually due to intermittent phonon scattering. Such behavior may be caused by non-linear interactions between phonon modes [18-19]. As is well-known, the perpetual motion of phonons is caused by the exchange of energy with the heat bath. This suggests that an optimal exchange of energy with the heat bath takes place if the intermittent phonon scattering exhibits extreme variance. Accordingly, 1/f noise is caused by the extreme property of phonon scattering.

$\zeta(\hat{\mu}_{on})$ is the normalization factor whereby $0.5 \leq \zeta(\hat{\mu}_{on}) \leq 0.6$ (see Appendix). Consequently, the average number of $M_t$-clusters is

$$A_{M_t} = M_t \hat{q}_{M_t} = \frac{\zeta(\hat{\mu}_{on})}{M_t^{\hat{\mu}_{on}(t)-1}}. \tag{40}$$

Using $\hat{\mu}_{on}(t)$ in Table 1 and Eq. (A.8), this is transformed to

$r > 1$:
$$A_{M_t} = \zeta(t)\frac{r}{M_t} = \zeta(t)\frac{\tau_{off}}{t}. \tag{41}$$

So far we have regarded only <u>one</u> realization of the IPP. Taking all X-centers in the probe volume into account, the average number of $M_t$-clusters is

$$A_X(t) = N_X A_{M_t}. \tag{42}$$

Substituting herein Eq. (41), we find

$$A_X(t) = N_X \zeta(t)\frac{r}{M_t} = N_X \zeta(t)\frac{\tau_{off}}{t} \tag{43}$$

decreasing with time $t$. The observation of a lowest frequency limit requires that at least <u>one</u> $M_t$-cluster occurs. Denoting the time fulfilling this condition by $T_X$ this is equivalent to

$$A_X(T_X) = 1. \tag{44}$$

For large $T_X$ and for moderate values of $r$, the normalization factor $\zeta(T_X) \approx 0.6$ (see Appendix) leading to

$$M_{T_X} = \frac{T_X}{\tau_X} = 0.6\, r N_X \tag{45}$$

and to a longest on-time by

$\tau_{off} > \tau_X$:
$$T_X = 0.6\, N_X \tau_{off}. \tag{46}$$

For $t > T_X$ the average number of $M_t$-clusters in the probe volume $A_X(t)$ rapidly fades out with increasing time $t$. Correspondingly, we find the onset of a plateau indicating the lowest frequency limit $f_X$. According to Eq. (18), this lowest frequency limit of 1/f noise in semiconductor materials is obtained by $f_X = 1/2M_{T_X}\tau_X$ resulting in

$\tau_{off} > \tau_X$:
$$f_X \approx \frac{0.8}{N_X \tau_{off}}. \tag{47}$$

For $t \gg T_X$ a distinct plateau appears below $f_X$ (see Fig. 5). Actually, due to fading out of $M_t$-clusters, the transition to the plateau is not abrupt as shown in Fig. 5 but smooth. Choosing $\tau_X = 10^{-5}$ s and $r = 10$, Table 2 summarizes $f_X$ and $T_X$ for several values of $N_X$.

For $r \leq 1$, the exponent $\hat{\mu}_{on} = 2$ and $\hat{q}_{M_t} = 0.6/M_t^2$. Repeating above steps we end up with

$\tau_{off} \leq \tau_X$:
$$T_X \approx 0.6\, N_X \tau_X \tag{48}$$

and

$\tau_{off} \leq \tau_X$:
$$f_X \approx \frac{0.8}{N_X \tau_X}. \tag{49}$$

Table 2. $f_X$, $T_X$ and $M_{T_X}$ for several values of $N_X$; the inter event time $\tau_X = 10^{-5}$ s and the normalized off-time $r = 10$.

| $N_X$ | $10^{15}$ | $10^{12}$ | $10^{9}$ |
|---|---|---|---|
| $f_X$ | $8 \cdot 10^{-12}$ Hz | $8 \cdot 10^{-9}$ Hz | $8 \cdot 10^{-6}$ Hz |
| $T_X$ | $6 \cdot 10^{10}$ s $\approx$ 1903 y | $6 \cdot 10^{7}$ s $\approx$ 694 d | $6 \cdot 10^{4}$ s $\approx$ 16,7 h |
| $M_{T_X}$ | $6 \cdot 10^{15}$ | $6 \cdot 10^{12}$ | $6 \cdot 10^{9}$ |

### 3.1. *The dependence of 1/f noise and g-r noise on time*

According to Eq. (A.2), a mean cluster size as a function of time is

$r > 1$:
$$\overline{\widehat{N}_c(t)} \approx \frac{6}{\pi^2}\left\{\frac{1-\hat{\mu}_{on}(t)}{2-\hat{\mu}_{on}(t)}\frac{M_t^{2-\hat{\mu}_{on}(t)}-1}{M_t^{1-\hat{\mu}_{on}(t)}-1} + C_E\right\}. \tag{50}$$

Substituting $\hat{\mu}_{on}(t)$ in Table 1, this is transformed to

$r > 1$:
$$\overline{\widehat{N}_c(t)} \approx \frac{6}{\pi^2}\left\{\frac{r-1}{\log r}\, \log\left(\frac{M_t}{r}\right) + C_E\right\} \tag{51}$$

increasing logarithmically with time $t$. For $r \leq 1$, a mean cluster size is

$r \leq 1$:
$$\overline{\widehat{N}_c(t)} \approx \frac{6}{\pi^2}\left\{\ln\left(\frac{t}{\tau_X}\right) + C_E\right\} \tag{52}$$

independent of $r$. As a consequence, also the on-time increases with time by

$$\hat{\tau}_{on}(t) = \overline{\widehat{N}_c(t)}\,\tau_X. \tag{53}$$

Substituting this into Eqs. (13) and (23), the pre-factors of 1/f noise and g-r noise are obtained by

$$\hat{\alpha}_{im}(t) = 2C(r)\frac{\tau_X}{\hat{\tau}_{on}(t)+\tau_{off}} = \frac{2C(r)}{\overline{N_c(t)}+r} \qquad (54)$$

and

$$\hat{\beta}_{im}(t) = \frac{\hat{\tau}_{on}(t)}{\hat{\tau}_{on}(t)+\tau_{off}} = \frac{\overline{N_c(t)}}{\overline{N_c(t)}+r}. \qquad (55)$$

Choosing $\tau_X = 10^{-5}$s, Figs. 8 and 9 show $\hat{\alpha}_{im}(t)$ and $\hat{\beta}_{im}(t)$ for several values of $T_X$ and of $r$. 1/f noise decreases and reduced g-r noise increases with time $t$ attaining constant values for $t \geq T_X$. For $T_X = 6 \cdot 10^{10}$ s, the transition of $\hat{\alpha}_{im}(t)$ and $\hat{\beta}_{im}(t)$ to constant values is beyond the scale of the t-axis (see Figs. 8.a and 9.a). For $T_X = 6 \cdot 10^4$ s this transition is within the scale of the t-axis (see Figs. 8.b and 9.b). The transition to a plateau, as is seen in these Figures, is not actually abrupt. Rather, it is a smooth transition because the average number of $M_t$-clusters gradually fades out as soon as $t > T_X$.

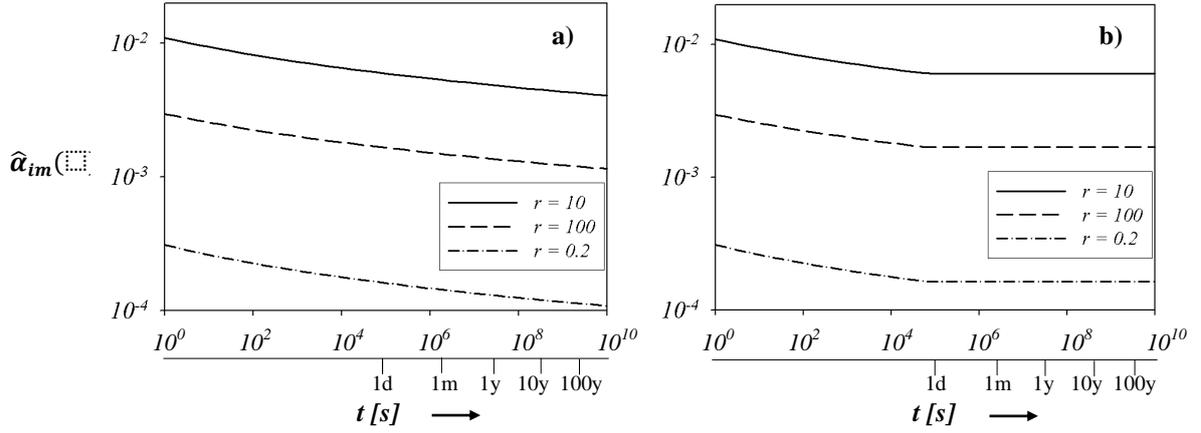

Fig. 8. The coefficient $\hat{\alpha}_{im}(t)$ versus time for several values of $r$. Fig. 8.a. For $T_X = 6 \cdot 10^{10}$ s, the transition to constant values is beyond the scale of the t-axis. Fig. 8.b. For $T_X = 6 \cdot 10^4$ s, this transition is within the scale of the t-axis. Herein, d = day, m = month and y = year.

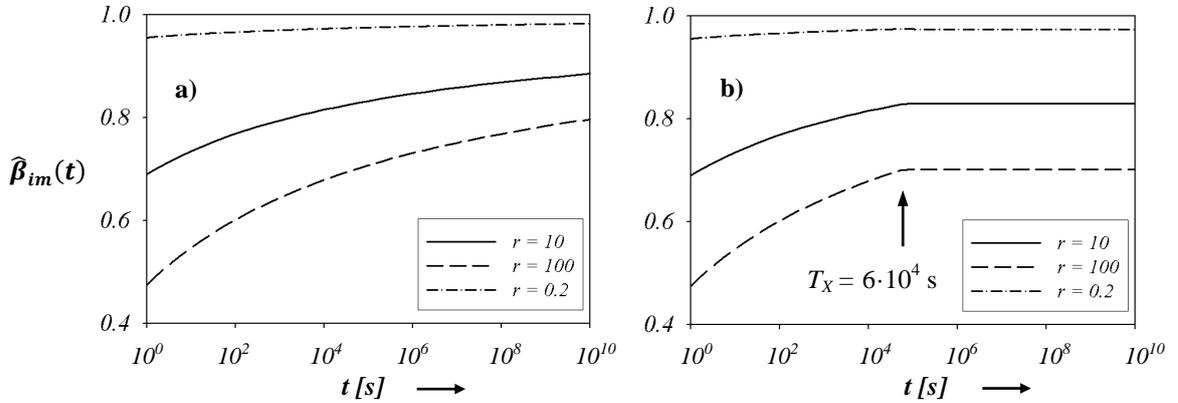

Fig. 9. The coefficient $\hat{\beta}_{im}(t)$ versus time for several values of $r$. Fig. 9.a. For $T_X = 6 \cdot 10^{10}$ s. Fig. 9.b. For $T_X = 6 \cdot 10^4$ s.

### 4. Results and Discussion

When measuring 1/f noise in semiconductors, we observe a 1/f shape which ends abruptly at the frequency $1/t$, where $t$ is the observation time [20]. An observation time of about 10 days corresponds to a lower frequency limit of about $10^{-6}$ Hz. A further extension of the observation time is a challenge not only to the patience of the experimenter, but also to the stability of instruments. A lowest frequency limit where the 1/f shape merges into a plateau has not yet been reported; we do not even know if such a lowest limit exists

In this paper, we report such a lowest limit of 1/f noise in semiconductor materials. Our considerations are based on an intermittent g-r process which is gated by off-times $\tau_{off}$. By analogy to blinking single quantum dots and other materials, the on-times are assumed to be power-law distributed as $1/t^{\mu_{on}}$. As a result, a generalized form of Hooge's relation is derived applying also for a slope $b \neq 1$. We also investigate the behavior of the slope $b$ and of the exponent $\mu_{on}$, both of which are controlled by $\tau_{off}$ (see Fig. 6). We find a slope $0.8 \leq b \leq 1.2$ corresponding to an exponent $1.7 \leq \mu_{on} \leq 2$. The slope $b$ and the exponent $\mu_{on}$ depend on the time $t$; here time $t$ is defined as the age of the semiconductor material after doping.

It is an essential feature of the intermittent g-r process that the power-law distribution of the on-times depends on time *t*. We estimate the average number of the longest on-times occurring within time *t*. The condition that at least <u>one</u> longest on-time $T_X$ occurs leads to $T_X = 0.6\, N_X \tau_{off}$; the corresponding lowest frequency limit is $f_X \approx 0.8/N_X \tau_{off}$. Below $f_X$ the 1/f shape merges into a plateau.

As an example, we choose $\tau_{off} = 10^{-4}$ s. For high doping ($N_X = 10^{15}$), we find $f_X \approx 10^{-11}$ Hz, which would lead to an observation time of about 1900 years. For low doping ($N_X = 10^9$) $f_X \approx 10^{-5}$ Hz corresponding to an observation time of about 17 hours, which is a reasonable time for finding a plateau appearing below $f_X$. This suggests that a lowest limit to 1/f noise in semiconductor materials can be found in choosing the number of X-centers $N_X$ as small as possible.

The time-dependent power-law distribution of the on-times also leads to time-dependent pre-factors for 1/f noise and g-r noise. After doping, a relaxation process takes place: 1/f noise decreases and reduced g-r noise increases with time *t* i.e. with the age of the semiconductor after doping. However, this relaxation process is rather slow. During the first day after high doping, 1/f noise decreases by a factor of about ½ (see Fig. 8a). A further decrease by a factor ½ takes about 100 years! Hence, a decrease in 1/f noise with time is best observed for "fresh" material (*t* < one day). Our considerations are in qualitative agreement with investigations on time-dependent processes due to anomalous diffusion, which also show ageing effects [21].

For *t* > $T_X$, the average number of the longest on-times fades out, and the relaxation process comes to an end. As a consequence, 1/f noise as a function of time merges into a plateau (see Fig. 8.b). Such behavior should be observed for low doping and for "old" material (*t* > 1day). Similar considerations also apply to the increase in g-r noise and changes in the slope *b* and exponent $\mu_{on}$ with time.

Assuming that a lowest frequency limit really exists, the total power of 1/f noise is finite, and the divergence problem of 1/f noise mentioned in the Introduction would vanish into thin air. The existence of a lowest limit is also a touchstone for an intermittent g-r process.

This research did not receive any specific grant from funding agencies in the public, commercial, or not-for-profit sectors.

**Mathematical Appendix.**

For $\mu_{on} = 2$, the mean number of spikes in a cluster (= first moment of cluster size) is obtained by

$$\overline{N_c} \approx \frac{6}{\pi^2}(\ln M_t + C_E) \tag{A.1}$$

where $C_E \approx 0.5772...$ is Euler's constant. For $\mu_{on} < 2$ the sum $\overline{N_c} = \sum n\, q_n$ is replaced by integration leading to

$$\overline{N_c} \approx \frac{6}{\pi^2}\left\{\frac{1-\mu_{on}}{2-\mu_{on}}\frac{M_t^{2-\mu_{on}}-1}{M_t^{1-\mu_{on}}-1} + C_E\right\}. \tag{A.2}$$

The factor $\frac{6}{\pi^2}$ and the addition of $C_E$ guarantee the equivalence to Eq. (A.1) for $\mu_{on} \to 2$. For $\mu_{on} = 2$, second moment of cluster size is

$$\overline{N_c^2} \approx \frac{6}{\pi^2} M_t \tag{A.3}$$

and by analogy to Eq. (A.2) for $\mu_{on} < 2$

$$\overline{N_c^2} \approx \frac{6}{\pi^2}\frac{1-\mu_{on}}{3-\mu_{on}}\frac{M_t^{3-\mu_{on}}-1}{M_t^{1-\mu_{on}}-1}. \tag{A.4}$$

According to Eq. (9), the cluster size distribution is

$n = 1, 2,...\, M_t$:
$$q_n = \frac{\zeta(\mu_{on})}{n^{\mu_{on}}}. \tag{A.5}$$

The normalization factor is a truncated Zeta function [15]

$$\zeta(\mu_{on}) = \sum_{n=1}^{M_t} \frac{1}{n^{\mu_{on}}}. \tag{A.6}$$

For $M_t \gg 1$, Table 3 shows the normalization factor for several values of $\mu_{on}$ with sufficient accuracy.

Table 3. The normalization factor $\zeta(\mu_{on})$ for several values of $\mu_{on}$.

| $\mu_{on} =$ | 1.7 | 1.8 | 1.9 | 2.0 |
|---|---|---|---|---|
| $\zeta(\mu_{on}) \approx$ | 0.487 | 0.532 | 0.572 | 0.608 |

A linear interpolation leads to

$$\zeta(\mu_{on}) \approx 0.4\, \mu_{on} - 0.2 \tag{A.7}$$

as a rough approximation in cases of practical interest. Substituting $\hat{\mu}_{on}(t)$ in Table 1, the normalization factor depends on time by

$$\zeta(t) \approx 0.6 - 0.4 \frac{\log(r)}{\log(M_t)} \tag{A.8}$$

applying for $1 < r < M_t$. For $r \leq 1$, the exponent $\hat{\mu}_{on}(t) = 2$ and the normalization factor is independent of time $\zeta(t) \approx 0.6$.